\documentclass[12pt,titlepage]{article}

\usepackage{html,hyperref}
\usepackage[pdftex]{graphicx}
\usepackage{multicol}

\begin{document}

\title{Nearby Galaxies: Templates for Galaxies Across Cosmic Time}

\author{F. J. Lockman\thanks{contact author: {\tt
      jlockman@nrao.edu}}, J. Ott \thanks{National Radio
      Astronomy Observatory, P.O. Box 2, 
Green Bank, WV 24944 // P.O.Box O, Socorro, NM, 87801}
}

\date{2009-02-15 \\[5ex]
\begin{minipage}[h]{6.5in}
  \begin{center}
  \includegraphics[width=4.0in]{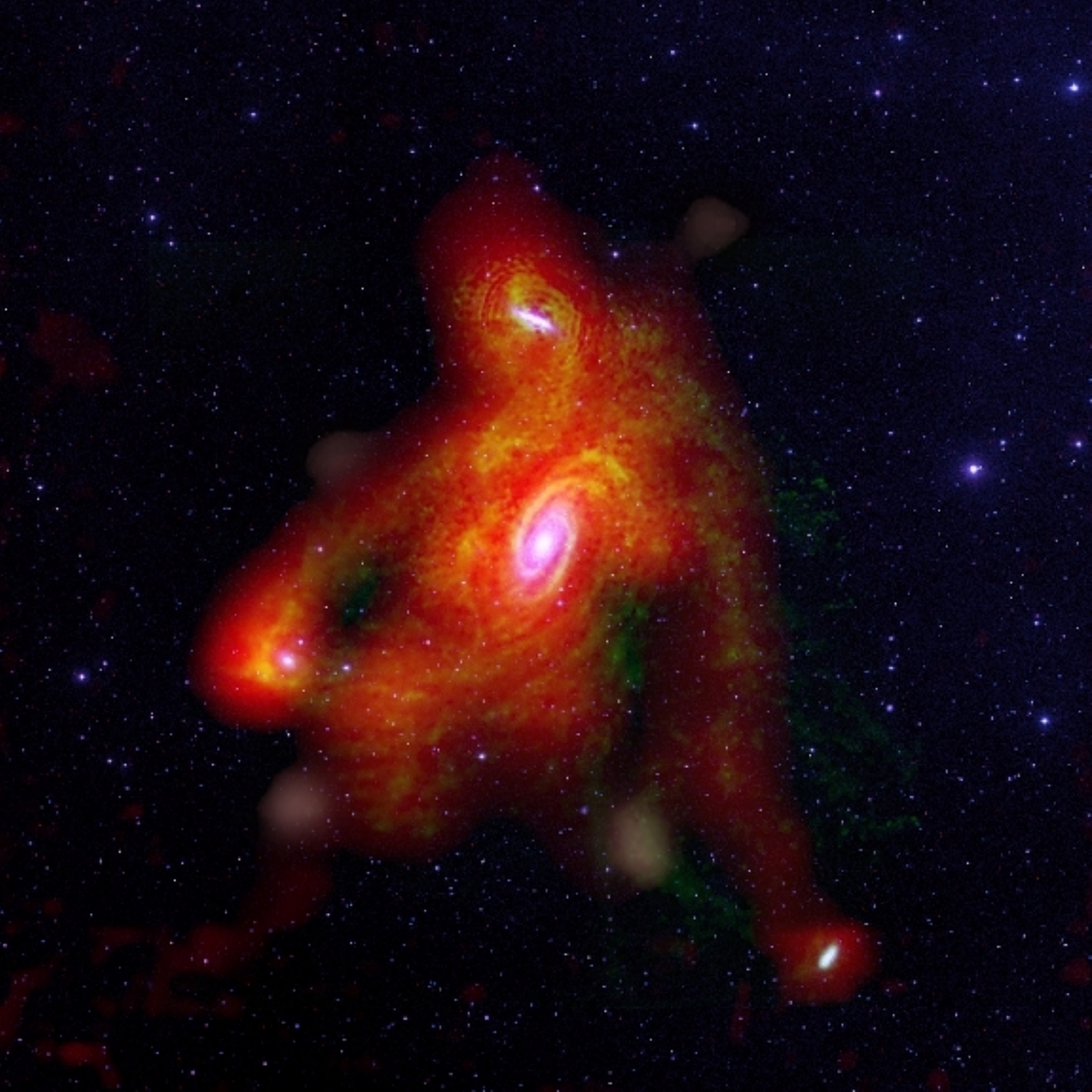}\\
  \end{center}
\end{minipage}
}

\maketitle

\begin{center}
{\Large \sc Nearby Galaxies: Templates for Galaxies 
 \break  Across Cosmic Time}
\end{center}

\begin{quote} {\bf Abstract:} Studies of nearby galaxies including the
  Milky Way have provided fundamental information on the evolution of
  structure in the Universe, the existence and nature of dark matter,
  the origin and evolution of galaxies, and the global features of
  star formation.  Yet despite decades of work, many of the most basic
  aspects of galaxies and their environments remain a mystery.  In
  this paper we describe some outstanding problems in this area and
  the ways in which large radio facilities will contribute to further
  progress.

\end{quote}


\section{Introduction}

The nearby Universe offers us the products of cosmic evolution --
galaxies --  the archetypes for all objects elsewhere in the
Universe.  The Milky Way (MW) and our closest neighbors like the
Magellanic Clouds provide the best data for understanding how a
galaxy's structure influences the ability of gas to form stars.  Local
galaxies can be used to derive star formation (SF) laws and to characterize
the broad range of galaxy types, from ellipticals to irregulars, from
spirals to active galaxies.  Locally we can examine galaxy
interactions and group and cluster influences.  We can study nearby
starbursts, HI supershells, and nuclear activity in detail.  A
knowledge of local galaxy parameters and their origin is critical to
understanding objects at large redshifts, where we will require
well-characterized templates of galaxies of all morphologies and
Hubble types derived from studies of the nearby Universe.

Some key questions that motivate the study of nearby galaxies include 

\begin{list}{$\bullet$}{
  \setlength{\topsep}{0ex}\setlength{\itemsep}{0ex plus0.2ex}
  \setlength{\parsep}{0.5ex plus0.2ex minus0.1ex}}
\item \textbf{How did galaxies grow?} 

\item \textbf{What connects galactic structure and star formation?}
   
\item \textbf{How do feedback processes affect galactic evolution?}

\item \textbf{What is the origin and importance of magnetic fields?}

\item \textbf{How are dark and baryonic matter related on the galactic scale?}
\end{list}

The image of the M81/M82 group shown in Fig.\,\ref{fig:comb}(left)
illustrates the range of phenomena accessible for study in the nearby
Universe: galaxy interactions, spiral structure, large-scale star
formation, molecular clouds, 
group and cluster dynamics, and starbursts.  In the coming
decade study of nearby galaxies will remain a vital activity.





\section{How do galaxies grow?}

Cold dark matter plus dark energy (``$\Lambda$CDM'') models predict
that galaxies are formed when baryons condense within dark matter
halos, which in turn are connected via thin filaments: the cosmic
web. The evolution of galaxies is driven by interactions and
mergers, accretion of fresh gas, and feedback from stellar evolution.
At every stage there are major gaps in our understanding.

Gas accretion, not only at early times but 
continuously to the present, appears to be indispensable given the 
chemistry and star formation history of today's
galaxies. In the local Universe, obvious gas accretion mechanisms include 
minor and major mergers of galaxies, galaxy interactions, and flows
from gas in galaxy halos or from the cosmic web.  Our best hope of
understanding the details of these processess is by observing them
locally.  By studying  galaxy groups we can determine the
rate and process of galaxy interactions
and can delineate the history and future of galaxy mergers,
e.g. the Magellanic Stream and the M81/M82 group HI trails.  In nearby
galaxies and the MW, high-velocity HI clouds may be evidence of
current cool gas accretion \cite{sancisi}.
Study of the cosmic web  has its own set of
questions: What is the metallicity of the web? Is there enough
material and can it be transferred to galaxies efficiently enough to
satisfy the chemistry model requirements? Does the local cosmic web
behave like high-z Ly-alpha absorbers? \cite{aracil,cote}

\section{What connects star formation to galaxy structure?}

\begin{figure}[t!]
  \includegraphics[width=\textwidth]{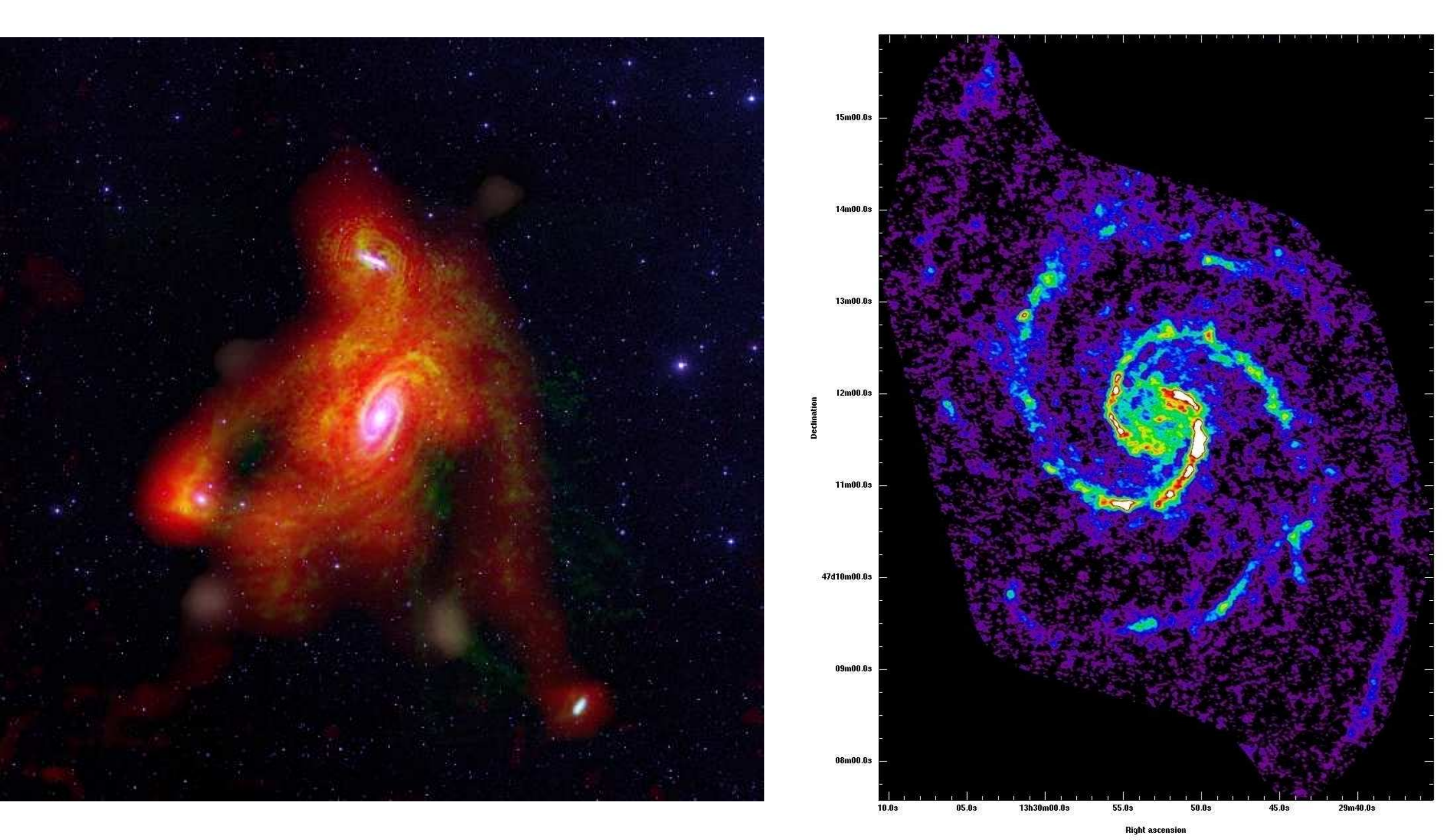}
  \caption{\label{fig:comb} 
 {\sc {\bf Left:}
  The M81/M82 group as observed in HI with the GBT and the VLA,
  superimposed on an optical image. This field contains a range of
  galaxy types including a starburst, evidence for strong
  interactions, and a fairly normal star-forming spiral \cite{chynoweth}.  
 {\bf Right:} CO(1-0) emission from the galaxy M51
  showing the spiral pattern of molecular clouds \cite{Koda07}.  
   }}
\end{figure}

Today, the brightest stars form in spiral arms of galaxies. This
simple observational result demonstrates the tight connection between
galactic dynamics, which is largely determined by dark matter, and
star formation. We know that compression and cooling of atomic gas
leads to the formation of molecular clouds, but there are many gaps in
our understanding of this process. On large scales,  molecular
clouds are embedded in the densest atomic gas reservoirs. On smaller
scales, however, the simple correlation breaks down and the two gas
phases are displaced from each other \cite{ott}. The location and mass
of molecular clouds is thus not simply related to the distribution of
the atomic phase; global as well as local effects are important for
the formation of molecular clouds (see Fig.\,\ref{fig:comb}[right] of
CO in M\,51).



The mass function of individual dense cores in a molecular cloud has a
striking similarity to the initial mass function of stars, implying
that the core mass function is somehow preserved throughout star
formation, in spite of the small fraction of gas that is converted
into stars (low star formation efficiency) and the very complex
processes that are involved in the gravitational collapse of gas into
a compact object \cite{smith,alves,padoan}. An integrated way to
express this property is the Kennicutt-Schmidt law. The study of local
and global star formation processes is one of the major themes in the
field of nearby galaxy research. Volume limited surveys of all
galaxies within a distance of $\sim $10 Mpc are currently underway at
many wavelengths (the Spitzer Local Volume Legacy [LVL] Survey, ANGST,
VLA-ANGST, (Little) THINGS, 11HUGS, ATLAS$^{\rm 3D}$, etc.), in a
program to build up a statistically meaningful sample 
useful in deriving star formation laws and evolutionary pathways. The
results of these surveys will provide the foundation for the
interpretation of observations at higher redshifts and for
understanding the evolution of star formation as a function of cosmic
time.

\section{Feedback processes in Galactic Evolution}

Galaxy growth is not a one-way route. The accretion of gas onto
galaxies provides the material for further star formation which
ultimately produces massive stars, supernovae, and black holes,
which in turn inject material and energy back into the surrounding 
ISM \cite{veilleux}.    Feedback processes are
indispensable elements of cosmic and galaxy evolution models along the
Hubble sequence, however, feedback is one of the most poorly
constrained parameters in galaxy models.  What governs the exchange of
matter between a galactic disk and the intergalactic medium?

The most prominent examples of feedback are certainly the bright
starburst and active galaxies such as M82 and Cen~A. Depending
on the gravitational potential of the host, however, lower level star
formation across galactic disks could be even as effective in
returning energy and metals to the IGM as nucleated starbursts. This
effect is amplified by the sheer number of low mass galaxies. Other
gas removal mechanisms include tidal interactions, ram pressure
stripping, and galaxy harassment. Those less obvious, but possibly very
important facets of feedback can be studied only in the nearby
Universe. To understand distant starbursts we need 
templates from the local Universe (e.g., M82) for understanding the
underlying mechanisms that drive, maintain, and terminate starburst
events.

%
%
%
%

\subsection{Feedback from Star Formation}

Energy and metals produced as the byproduct of stellar evolution are
circulated throughout galaxies and into the intergalactic medium
through phenomena such as jets, winds, superbubbles, and galactic
fountains. The redistribution of matter and energy affects galactic
and stellar evolution and may even modify the dark matter halo
\cite{lagos,pedrosa}. A starburst can be strong enough to remove all
gas thus stopping further SF processes. On the other hand, density
enhancements around SF regions may trigger further, secondary SF. A
key question is if star formation must be initiated by
large, galactic scale processes such as spiral arms and bars, or
whether the turbulent ISM itself might trigger widespread,
low-level SF.
Episodic star formation is observed in many galaxies, often triggered
by interactions.  But is an interaction always required, or does the
cooling of heated-up gas from former SF events determine the timescale
for subsequent bursts and thus the episode timescales?

Each SF event injects large amounts of energy into the surrounding
medium. Yet there are regions characterized by very high levels of
star formation in a relatively small volume.  How long can this be
maintained? How are star-forming regions fed with gas without
destroying them? Why are there large amounts of molecular gas adjacent
to a region of extremely large ionizing flux, as occurs in starbursts?
And finally, if starbursts can be maintained over a long period, how
are they stopped? Is it pure starvation of feeding material or are
there other mechanisms in place? Is there the equivalent of the
Eddington luminosity limit for starbursts in nearby galaxies?

\subsection{Feedback from Galactic Nuclei}

During the last decade, an intriguing, almost linear, relationship was
extablished between the bulge mass of galaxies and the mass of their
central black hole \cite{magorrian}.  This is surprising as the
bulge is $\sim700$ times more massive than the central black
hole and should dominate the system.  Feedback processes 
from the black hole, however, might affect star
formation in its surroundings and thus the bulge mass
\cite{rafferty}.  This feedback 
can be studied in great detail only in nearby galaxies. Active
nuclei can produce large scale jets that influence 
 interstellar and intergalactic space to Mpc scales. Jets
may induce star formation or prevent star formation  
 \cite{dopita,lagos}.  In addition,
AGN are frequently observed in conjunction with nucleated starbursts,
sites of the most vigorous star formation activities in the
Universe. The bi-conical winds produced by strong nuclear starbursts
have a wider opening angle than jets and are able to remove larger
quantities of gas, and importantly, carry freshly produced metals 
out of a galaxy.   To understand the
feedback from galactic nuclei and to study their influence on their
environments in detail, it is indispensable to study nearby objects
such as Cen~A, M\,82, NGC\,253, Fornax\,A, or M\,87 at high
angular resolution and with sufficient sensitivity. In
particular, the MW central black hole, Sgr A*, provides a unique
window into the connection between the massive central
object and its immediate surroundings. 


\section{Molecules and Chemistry in the Local Universe}

One of the major themes of radio astronomy has been the study of the
interstellar medium (ISM), which breaks down into various phases at
different temperatures, densities and ionization fractions.  The
chemical processes in the ISM produce a rich set of organic molecules,
exotic species, and ``pre-biotic'' molecules that may be relevant to
the origin of life on earth.  For several decades molecules have been
used as probes of intersteller processess, transforming 
 our understanding of star formation and
evolution, and the 
physical conditions in molecular clouds (e.g., \cite{iau231}).

The coming decade will see increased activity in this area as well as 
a move to use astronomical observations to study chemistry itself.   
Terrestrial laboratories  are largely limited to studing reactions 
in liquids or high-density gases.  A question such as: 
{\it how does non-equilibrium chemistry proceed in a weakly ionized 
gas in the presence of magnetic fields?}   crosses 
traditional disciplinary boundries as it can be answered only by 
astronomical observations in conjunction with theoretical and 
laboratory studies.  

Radio astronomy is a unique tool for fundamental chemistry. The lowest
rotational transitions of molecules are in the centimeter to
sub-millimeter wavelength range.  The lines are weak and often from
extended sources, so progress thusfar has been limited to study of
only a handful of molecular clouds, almost entirely in the MW.  In
coming years, however, studies will be extended throughout the MW and
to nearby galaxies as specific chemistry questions are asked that
require probes of  specific physical conditions.


\section{The Origin and Importance of Magnetic Fields}

Magnetic fields affect gas motions in galaxies and influence, perhaps
control, the collapse of clouds into stars.  They couple energy from
supernovae to the interstellar medium.  They control the density and
distribution of cosmic rays.  Yet the origin of the fields, their
evolution, and many critical aspects of their interaction with gas
remain controversial \cite{ZweibelHeiles,kulsrudzweibel,beck}.  How
are magnetic fields generated in galaxies?  Is the field related to
structure and dynamics of a galaxy? How do the magnetic fields
(re)distribute cosmic rays and halo gas?  Does reconnection provide an
important heat source for a galaxy's disk and halo?  How does the
field evolve as a dense core contracts to become a star?

At the present time there are only a few galaxies that have good field
measurements.  What is needed is a much larger sample so that we can
understand how magnetic structures derive from other galaxy
properties.  Zeeman splitting can provide the strength of the magnetic
field in gas-rich regions of galaxies. Rotation measures of
background sources will probe many sightlines and thus 
supply points in a
grid on which to fit on large scale magnetic field models. Radio
polarization sheds light on  SN and SF dominated regions
and events like galactic outflows.

\section{Dark Matter}

About one-quarter of the mass of the Universe is in the form of Dark
Matter (DM), likely consisting of cold, non-baryonic particles.  
Outstanding questions include: What is the relationship between
the baryonic and dark matter components?  Are there ``naked"
dark-matter structures in galactic halos?  Do nearby galaxies have the
detailed structure implied by CDM models and where are the missing
satellite galaxies CDM predicts?  Are galactic warps coupled to the
dark matter?  What is the history of galaxy interactions in groups?
What is the distribution of mass in the local group and what is the
fate of the individual galaxies? A better characterization of the
amount of DM on different size scales and how it is distributed is a
critical need.

Astronomical observations can give considerable insight into the
nature of DM: through precise astrometry we can determine the motion
of nearby objects and deduce the local gravitational potential. Gas
motions in the Milky Way and other galaxies are sensitive to the
potential on many size scales. We can trace the movements of large
objects like galaxies in clusters. Studies of the structure of nearby
galaxies have already confronted DM models with significant constraints
\cite{deblok}.  This is an important and continuing activity.

\section{Fundamental Science Opportunities}

\begin{list}{$\bullet$}{
  \setlength{\topsep}{0ex}\setlength{\itemsep}{0ex plus0.2ex}
  \setlength{\parsep}{0.5ex plus0.2ex minus0.1ex}}
\item \textbf{Extending ``Galactic Astronomy'' to a variety of galaxies}
\end{list}

With increased collecting area and spatial resolution, topics 
 in ``Galactic astronomy'' will  be
studied in other galaxies. Subjects include the abundance and
distribution of halo gas and HVCs; the details of star formation on
the scale of individual pre-stellar cores; 
the relationship between atomic and molecular gas in SF regions; 
 the distribution of objects like stars, pulsars, SNe and 
planetary nebulae, and their 
influence on the surrounding ISM. 
 Currently this kind of science is restricted to
the Milky Way (with some data from our closest neighbors like the
Magellanic clouds) and is thus coupled to a specific type of galaxy
with its very own evolutionary status and history. To perform these
studies in a variety of galaxies -- from dwarf irregulars to massive
ellipticals, from isolated to interacting galaxies, from starbursting
to `red and dead' objects, from face-on to edge-on galaxies, from
barred to AGN environments -- will change our knowledge
fundamentally. 
Some work on this has been done already, as shown in the
figure of the M81/M82 group, and with some recent surveys like
THINGS \cite{walter}, but in combination with multi-wavelength studies
in the optical, near-UV and infrared, the next generation of radio
studies can characterize individual components of star formation in
galaxies of very different morphology, size, and evolutionary state. 
 Freed from observational restrictions that have confined
much of our studies to the immediate Galactic neighborhood, we will
derive vastly more accurate statistics on every aspect of 
star formation, galactic structure, and evolution.

\begin{list}{$\bullet$}{
  \setlength{\topsep}{0ex}\setlength{\itemsep}{0ex plus0.2ex}
  \setlength{\parsep}{0.5ex plus0.2ex minus0.1ex}}
\item \textbf{The Chemistry of the Local Universe}
\end{list}%

With increased sensitivity, radio 
astronomical observations 
will expand our understanding of basic chemical processes 
as they proceed under conditions not achievable in terrestrial 
laboratories,  to yield fundamental information on the nature 
of the chemical bond.  
The use of chemistry as an astrophysical probe, and astronomy 
as a tool for chemistry, will provide unique opportunities for 
advancement of both fields.

\section{Radio Instrumentation and Nearby Galaxies}

Radio observations are unique in providing information on galactic gas
in its ionized, neutral atomic and molecular phases over a range of
conditions and angular scales.  Radio observations are also unique in
their ability to measure magnetic fields through Faraday rotation, the
polarization of synchrotron emission, and the Zeeman effect in
spectral lines. Only radio observations have the extremely high angular
resolution necessary for the most precise astrometry.  In addition,
some of the most interesting regions (Infrared-dark clouds, galactic
nuclei) are deeply obscured at optical and infrared wavelengths, and
only radio observations provide access to the physics within.

\begin{list}{$\bullet$}{
  \setlength{\topsep}{0ex}\setlength{\itemsep}{0ex plus0.2ex}
  \setlength{\parsep}{0.5ex plus0.2ex minus0.1ex}}
\item \textbf{Radio capabilities provided in the next decade}
\end{list}

EVLA\footnote{Expanded Very Large Array {\tt
    http://www.aoc.nrao.edu/evla}} and GBT\footnote{Robert C. Byrd
  Green Bank Telescope {\tt http://www.gb.nrao.edu/gbt}} observations
of HI absorption against  background  sources may
reveal cosmic web filaments in absorption
along the line of sight, and, through Zeeman measurements, their 
magnetic field.  GBT and EVLA studies of HI
emission reveal the origin of HVCs, and their relation to Galactic
structure, in the Milky Way and nearby Galaxies. Measurement of HI in
fields around individual galaxies and in galaxy groups with the GBT
and EVLA reveal signs of outflows and interactions and their history.
The VLBA\footnote{Very Large Baseline Array {\tt http://www.vlba.nrao.edu/}} is being used to derive Galactic structure to unprecedented
accuracy, revising the mass of the Milky Way, and maping the 3D
velocity field of the local group through observations of proper
motions of galaxies.

EVLA, GBT and ALMA can map the molecular clouds and cold dust in
galaxies and galactic nucleii and determine their physical properties
and kinematics. 
They can study chemical processes under conditions in the ISM 
that can not be replicated on Earth.  
The GBT is the only instrument that can observe the
important CO(1-0) molecular line at essentially every redshift.
Detailed studies of Sgr A* with the EVLA and VLBA give information
that can be applied to other nuclei. EVLA and GBT can measure
magnetic fields in HI and OH through the Zeeman effect.

\begin{list}{$\bullet$}{
  \setlength{\topsep}{0ex}\setlength{\itemsep}{0ex plus0.2ex}
  \setlength{\parsep}{0.5ex plus0.2ex minus0.1ex}}
\item \textbf{Enhancements needed to achieve the science goals}
\end{list}

 The ultra-compact EVLA E array will provide sensitivity on
angular scales needed to make initial studies of the cosmic web in
21cm or radio continuum emission.  GBT focal plane cameras in the HI
line will allow deep mapping of wide areas, and at 22~GHz 
 will identify extragalactic water masers for local group proper
motion studies.  Focal plane cameras on the GBT will provide
wide-field, high-sensitivity mapping of molecular clouds in many 
molecular lines at wavelengths as short as 3mm, providing great
synergy with ALMA.   
EVLA E array will improve sensitivity to molecular emission on arc-min
scales.  Increased sensitivity of the VLBA and HSA will allow precise
astrometry on faint objects, and more precise distances to star-forming
regions.


\begin{multicols}{2}
\renewcommand{\baselinestretch}{0.65}

\end{multicols}
\end{document}